\documentclass[useAMS,usenatbib,12pt,preprint]{aastex}
\usepackage{amsmath}
\usepackage{amsfonts}
\usepackage{graphicx}
\usepackage{verbatim}
\usepackage{color}
\usepackage{aas_macros}

\begin{document}

\title{Order statistics of the early-type galaxy luminosity function}


\author{L\'aszl\'o Dobos\altaffilmark{1}, Istv\'an Csabai\altaffilmark{1}}
\altaffiltext{1}{E\"{o}tv\"{o}s Lor\'{a}nd University, Department of Physics of Complex Systems, Budapest, Hungary}
\email{dobos@complex.elte.hu}

\label{firstpage}

\begin{abstract}
We apply order statistics (OS) to the bright end ($M_r < -22$) of the luminosity distribution of early-type galaxies spectroscopically identified in the SDSS DR7 catalog. We show that an overall normalized luminosity function can be derived from the data that describes the distribution of red elliptical galaxies sufficiently for the purposes of OS in a broad redshift range of $0.1 \leq z \leq  0.5$. We calculate the typical OS quantities of this distribution numerically, measuring the expectation value and variance of the $k^{th}$ most luminous galaxy in a sample with cardinality $N$ over a large ensemble of such samples. From these statistical quantities we explain why and in what limit the $k^{th}$ most luminous galaxies can be used as standard candles for cosmological studies.

As a sample application of OS we show that galaxy counts in different redshift ranges can be easily estimated if the absolute magnitude of the few most luminous galaxies and the overall shape of the luminosity function is known (and does not change significantly with $z$). First we demonstrate that the absolute magnitude of the $k^{th}$ most luminous early-type galaxies can be estimated from galaxy number counts in the investigated redshift range. By reversing the method, galaxy counts can also be very easily calculated, for example, in redshift ranges where spectroscopic data is available only for the brightest sources.

Since our sample contains all bright galaxies including the brightest cluster galaxies (BCG), based on OS we argue that BCGs  can be considered as statistical extremes of a well-established Schechter luminosity distribution when galaxies are binned by redshift and not cluster-by-cluster. We presume that the reason behind this might be that luminous red ellipticals in galaxy clusters are \em not random \em samples of an overall luminosity distribution but biased by the fact that they are in a cluster containing the BCG. We show that a simple statistical toy model can reproduce the well-known magnitude gap between the BCG and the second brightest galaxy of the clusters.
\end{abstract}



\section{Introduction}

The physical properties of luminous red galaxies (LRG) make them suitable to select a consistently evolving sample in a broad redshift range. LRGs are the most massive stellar populations in the Universe which makes them a good proxy of matter distribution to approximately redshift $z \simeq 1.2$ with current observational techniques \citep{sdss, combo17, deep2, vvds, 2slaq, dr7}. Thanks to the significant break in their spectrum at 4000\,\AA \ it is relatively easy to determine their redshift even from spectra observed at low S/N ratio \citep{Eisenstein2001}. LRGs are thought to be assembled before or around $z \geq  2$ and the significant part of their stellar population has been passively evolving since then \citep{GunnOke1975, Ellis1997, AragonSalamanca1998, vanDokkum1998, Stanford1998, Burke2000, Wake2006, Maraston2009}. We used a consistent sample of luminous red galaxies taken from the Sloan Digital Sky Survey DR7 to demonstrate the statistical methods subject to this paper.

Order statistics (OS; in the special case of $k=1$ first order statistics or extreme value statistics -- EVS) is a field of statistics concentrating on determining expectation values and moments of extrema and the $k^{th}$ largest (smallest) value of samples of a given cardinality drawn from an underlying probability density distribution. Expectation values and higher moments of statistical quantities can be expressed as functions of the sample size $N$ and the order $k$. While these calculations are complex analytically, numerical integration of the formulae is very simple by the Monte-Carlo method which is essentially equivalent of computing the statistics of mock catalogs generated by using a predefined luminosity function. Extreme value theory applied to independent and identically distributed or weakly correlated variables yields several theorems about limiting distributions of extrema: the behaviour of the tail of the underlying distribution determines the kind of the distribution of extreme values and only a few of these exists as shown by \citet{FisherTippett1928, Gnedenko1948}.

The brightest LRGs can be found in the central regions of galaxy groups and clusters and have been successfully used as standard candles because of the small scatter in their intrinsic luminosity, once correction for their evolution is applied \citep{Sandage1972, Sandage1976, Postman1995, Whiley2008}. It is a widely discussed question whether these galaxies can be considered statistical extremes of the same luminosity distribution as the rest of the early-type galaxies or they belong to a different distribution \citep{Geller1976, Geller1983, Loh2006, DeLucia2007, vonderLinden2007, Liu2008, Lin2010}. The latter case would suggest certain selection effects and different physical processes in their formation and evolution. One of the main statistical arguments against that they belong to the same luminosity distribution as the rest of early-type galaxies is the magnitude gap of $\Delta M \simeq 0.8 \,\,\text{mag}$ between the BCGs and the second brightest galaxies of the same clusters \citep{Loh2006}.

Earlier astrostatistical studies applied EVS to investigate the distribution of the brightest cluster galaxies and claimed that the distribution of the brightest \em cluster \em galaxies is different from those of brightest \em group \em galaxies. It was stated that the luminosity distribution of the latter follows the extreme value distribution of an underlying distribution with tail extending to infinity but decaying faster than any power law \citep{Bhavsar1985, Bernstein2001}. They did not take into account however, that for small values of $N$ significant corrections apply to the theoretical limiting extreme value distributions as shown by \citet{Gyorgyi2008, Manu2010}.

The structure of this paper is as follows. In Sec.~\ref{sec:data} we describe the dataset we used and our method to calculate the absolute magnitudes. In Sec.~\ref{sec:lumf} we determine the luminosity distribution of our sample. We give a short introduction to the basics of the area of extreme value and order statistics in Sec.~\ref{sec:evs} and calculate the most interesting OS quantities of the luminosity distribution of our sample. In Sec.~\ref{sec:counts} we demonstrate the applicability of OS to determine galaxy counts at high redshifts where only the brightest galaxies were observed. In Sec.~\ref{sec:bcg} we describe a very simple model that can reproduce the megnitude gap between the BCG and the second brightest galaxy of clusters. In Sec.~\ref{sec:discussion} we discuss the capabilities, limitations and relevance of EVS/OS in cosmological investigations including the {\it quest for the} brightest cluster galaxies.

Throughout this paper we adopt a flat $\Lambda$CDM cosmology with $h = 0.7$, $\Omega_M = 0.3$ and $\Omega_\Lambda = 0.7$. Where not otherwise noted, absolute magnitudes are considered K- and evolution-corrected and calculated from extinction corrected best radial fit model magnitudes.

\section{The data}
\label{sec:data}

As we based our sample selection on a stellar population evolution model and K- and evolution-corrected absolute magnitudes it is important to discuss this model before describing the selection criteria.

\subsection{K-correction and evolution correction}
\label{sec:kcorr}

\begin{figure}
    \begin{center}
      \input{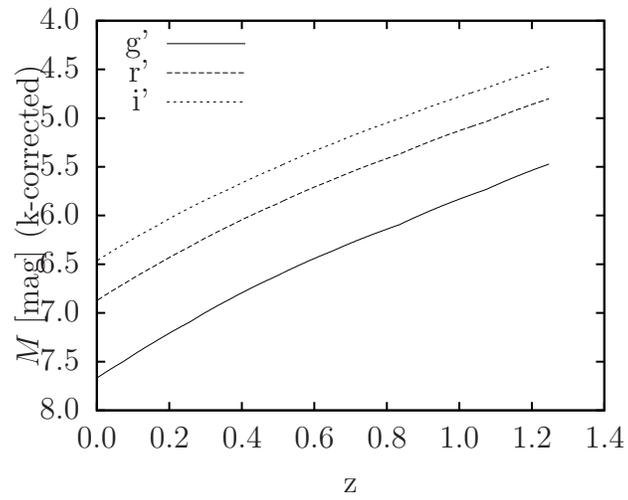}
      \caption{K-corrected absolute magnitudes (SDSS g', r', i' band, from bottom to top) of a passively evolving model of an early-type stellar population with an age of 12\,Gyr at $z = 0$, normalized to $1\,M_\odot$. The exponential change of luminosity of early-type galaxies with age is clearly visible and the effect has to be corrected if one wants to compare galaxies based on their absolute magnitudes. (Note the inverted scale.)}
      \label{fig:marmag}
	\end{center}
\end{figure}

Since the mass to light ratio of passively evolving stellar populations changes heavily with time it is very important to correct for stellar evolution when comparing galaxy counts at different epochs based on their measured magnitudes \citep{Eisenstein2001, LohPhD, Loh2006, Wake2006}. To correct for stellar evolution and determine a comparable absolute magnitudes we adopted the best-fit composite stellar population model of \citet{Maraston2009}. This semi-empirical model describes the evolution of the SDSS $g - r$ and $r - i$ colors of red sequence galaxies fairly well in the required redshift range. The model assumes that the stellar populations of the red sequence galaxies would be equally 12\,Gyr old at redshift $z = 0$ and they have been evolving passively since they had assembled at redshift $z \geq 2$. The original model spectra were computed in 1\,Gyr increments, we used linear interpolation between those time steps. Fig.~\ref{fig:marmag} shows the evolution of a K-corrected (i.e. not redshifted) passively evolving stellar population in the SDSS g', r', i' bands.

Following the recipe in \citet{Maraston2009} we computed the value of K-correction for the $u$, $g$, $r$, $i$ and $z$ filters of SDSS using the transmission curves \citep{Fukugita1996} from the SDSS web site for every galaxy as if they were at redshift $z = 0$. For evolution correction, we computed the flux ratios of a model at the observed redshift with age $t_1 = 12\,\text{Gyr} - t_{lb}$ and a model at $z=0$ with age $t_2 = 12\,\text{Gyr}$, where $t_{lb}$ is the cosmic look-back time.

Adopting a constantly evolving model that does not allow for scatter in the age of stellar populations at a given redshift seems to be too simplified but it turned out to work better than fitting models of varying age to the individual galaxies. Former studies focusing on early-type galaxies concluded that the dispersion in the age of high-mass ellipticals is in the $100 \,\, \text{Myr}$ regime \citep{Gallazzi2006}. Also, the scatter in K + e corrected absolute magnitudes is significantly higher when each galaxy's age is fitted individually. This is thought to be due to the degeneracy in photometric error scatter, and scattering due to age and metallicity around the magnitudes computed from the model spectra. Thus, the constantly evolving model works well for the vast majority of the brightest galaxies with passively evolving stellar populations. Also, galaxies experiencing episodic quiet star formation due to minor merging or smooth accretion are thought to have very similar optical colors as their completely passive counterparts \citep{Kaviraj2008}. The model cannot be applied to those galaxies that recently evolved onto the red sequence via merging or quenching of star formation from the blue sequence \citep{Bell2004, Bell2007}. However, the latter galaxies are thought to have the luminosity below our absolute magnitude limits.

\subsection{Notes on the evolution correction}
\label{sec:evo}

\begin{figure}
	\begin{center}
       \input{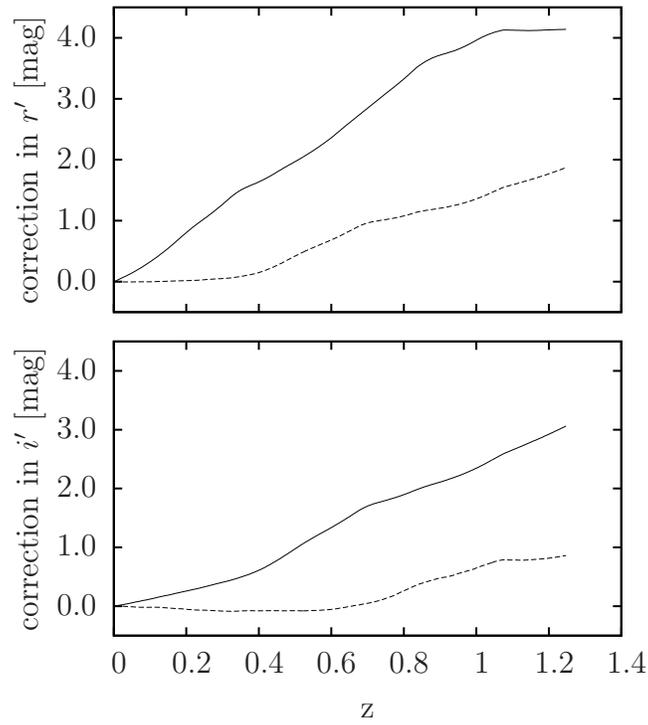}
       \caption{K-correction (solid line) and K+evolution correction (dashed line) in the SDSS r' band (top panel) and i' band (bottom panel) for a passively evolving early-type galaxy with an age of 12\,Gyr at $z = 0$. Note that the K+e correction is almost zero out to $z = 0.35$ and $z = 0.65$, respectively.}
       \label{fig:kcorr}
	\end{center}
\end{figure}

We would like to emphasize the importance of right evolution correction of the magnitudes of early-type galaxies in order to obtain physically meaningful results. As we observe galaxies at higher and higher redshifts, their colors become redder due to the shifting of the observed wavelength of the 4000 \AA\ break. On the other hand, as we look further out we see younger stellar populations that tend to be intrinsically bluer and brighter. These two effects compete in the SDSS $r$ and $i$ bands and they almost cancel out each other between $z = 0$ -- $0.35$ and $z = 0$ -- $0.6$ respectively, as can be seen in Fig.~\ref{fig:kcorr}. The difference between pure K-correction and K+e-correction can be as big as 0.5 mag which could distort the absolute magnitude limited co-moving density estimates significantly. 

Forthcoming large photometric surveys like PanSTARRS and LSST will not measure spectroscopic redshifts. Hence, reliable photometric redshift estimation of strongly evolving galaxies will be crucial in the future \citep{Padmanabhan2005}. Many current, template-based, widely used photometric redshift estimation algorithms do not take stellar population evolution into account directly \citep{Hyperz, Blanton2003, Blanton2007, Hildebrandt2010}. Certain algorithms solve this problem by using a numerous set of LRG templates with varying ages and attempt to fit the age from photometry alone without making an assumption on the covariance of the age and redshift. This is not a good practice in case of LRGs as photometric scatter and scatter from age and metallicity differences cannot be easily separated from optical broadband photometry. Also, using too many templates introduces too many degrees of freedom in fitting.

Empirical or semi-empirical template-based algorithms however are able to achieve good photometric redshift estimates by "fixing" the spectrum templates, by fitting them to the data, producing slightly distorted templates that have some sort of \em{effective evolution}\em ~encoded into the overall curve of the spectrum \citep{Csabai2000, Budavari2000, Blanton2003}. 

\subsection{Our subsample of SDSS DR7 spectro}
\label{sec:spectro}

For this study we used the SDSS DR7 spectroscopic data set, \texttt{spectro} hereafter \citep{dr7}. To create our early-type galaxy sample we selected all galaxies from \texttt{spectro} which had the best cross-correlation redshift confidence (but the minimum of $z_{conf} \geq  0.9$) with the LRG template of the SDSS \texttt{spectro} pipeline \citep{Eisenstein2001}. Since fibres of the SDSS spectrograph cover only the central 3" of extended sources, encompassing only the bulge of nearby spirals, low-redshift late-type spectra can closely resemble those of early-type galaxies. Though bulges of spirals are much fainter than LRGs, they can affect number counts calculated with lower magnitude limits. This issue can be partially resolved by taking observed photometric magnitudes and radial profiles into account which help to discriminate between bulges of spirals and elliptical galaxies. Consequently, we also required that the selected galaxies must be close enough to the trajectory defined by the evolving fiducial LRG model of \citet{Maraston2009} in the $g-r$; $r-i$; $i-z$ color-color space. The following constraint was set.

\begin{multline}
	\label{eq:colorsel}
	\frac{1}{9}((g - r) - (g_m - r_m)) ^ 2 + ((r - i) - (r_m - i_m)) ^ 2 + \\
	+ ((i - z) - (i_m - z_m)) ^ 2 < 0.015,
\end{multline}
where $g$, $r$, $i$ and $z$ are the observed, dereddened best-fitting model magnitudes, and $g_m$, $r_m$, $i_m$ and $z_m$ are the synthetic magnitudes of the stellar population model along the trajectory (refer to Sec.~\ref{sec:kcorr} on the model used).

Also because of the 3" spectroscopic fibre radius, additional constraints must have been set to restrict source selection to elliptical galaxies only. We required that all galaxies must have good de Vaucouleurs' radial profile fits in the r band (fraction of the de Vaucouleurs' function in the total radial profile fit is over 80\%, refer to \citet{SDSSEDR} for details on surface brightness fitting in SDSS) and that all galaxies have reliable redshift measurements (with over 90\% confidence).

We verified the most luminous 1,000 galaxies of the automatically selected sample by eye and excluded the outliers from the data set. The exclusion rate was about 1.5\% and the reason of exclusion usually was the closeness of a very bright object or "ghost" which confused the SDSS photo pipeline leading to invalid magnitudes. The selection process resulted in a sample of 142,762 galaxies ranging to $z = 0.63$ in redshift and covering 8,032 square degrees of the sky. The brightest part of the sample ($M_r < -22$) contains 73,459 galaxies. We estimated the \texttt{spectro} completeness factor as 0.9 in the density calculations.

The uniformity of our sample is not perfect below $z < 0.07$ due to the well-known properties of the SDSS: The SDSS photometric reduction algorithm might erroneously deblend galaxies with very large apparent diameter into two when more than one high surface brightness peak is detected within the same source. As we are interested in the biggest galaxies this issue can significantly effect the number counts at very low redshifts.  The total LRG sample is volume-limited to $z \ = 0.38$ \citep{Loh2006} but all galaxies brighter than $M_r \leq -22$ were observed up to redshift $z = 0.45$, galaxies brighter than $M_r \leq -22.5$ up to $z = 0.51$.

\section{The LRG luminousity function}
\label{sec:lumf}

In this section we estimate the magnitude distribution of our LRG sample and fit the parameters of the Schechter function in the form of

\begin{multline}
	\phi \left( M \right) \, \text{d}M = \phi_0 \exp \left\{ C (\alpha + 1) (M - M^*) - \right.\\
	\left. - \exp{ \left[ C (M - M^*) \right] } \right\} \, \text{d}M,
	\label{eq:schechter}
\end{multline}
where $C = -0.4 \log{10}$.

\begin{figure}
    \begin{center}
	  \hspace{24pt}
      \input{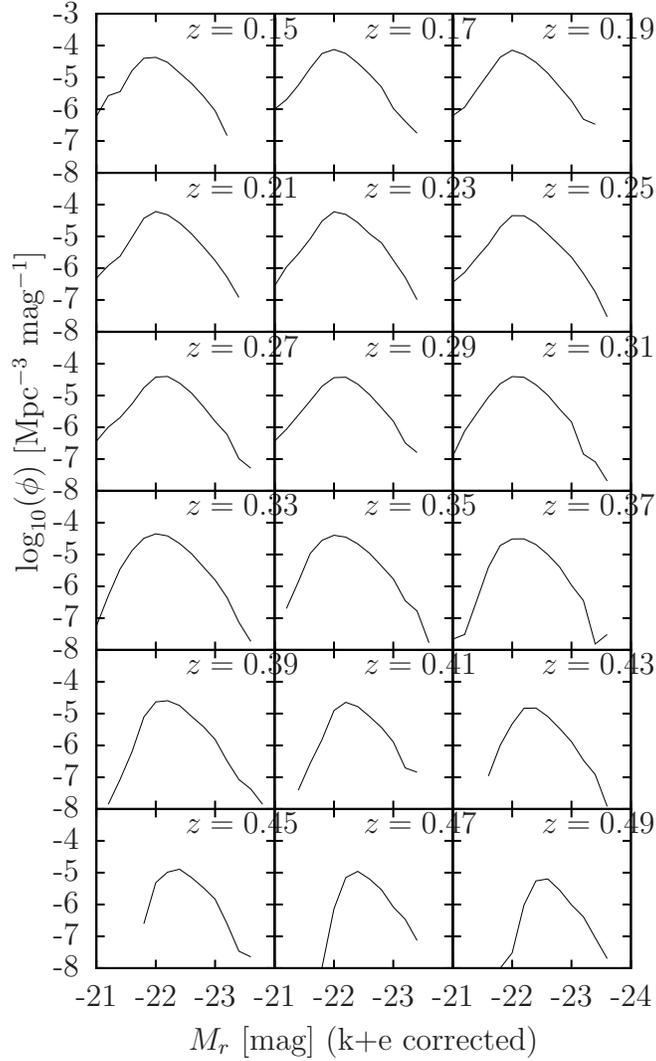}
      \vspace{32pt}
      \caption{Luminosity distribution of the early-type galaxies in redshift bins of $\Delta z = 0.02$ between $0.14 \leq z \leq 0.50$. The central redshift of the bins is indicated in the top right corner of each plot. The bright end of the sample ($M_r \leq  -22.5$) is volume-limited in the whole redshift range. The low luminosity turn-offs are due to the limiting magnitude of $m_r \leq  19.5$ of the spectroscopic targeting algorithm of SDSS. See Fig.~\ref{fig:lumf} for the same plot with the curves plotted over each other.}
      \label{fig:lumfbyz}
    \end{center}
\end{figure}

\begin{figure}
    \begin{center}
      \input{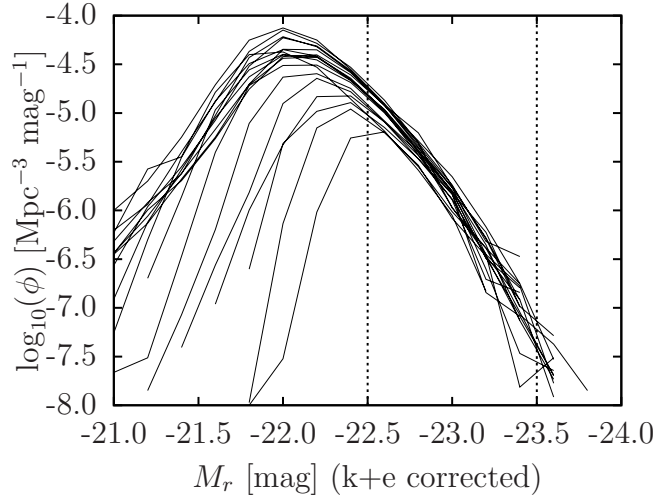}
      \caption{Luminosity distribution of the early-type galaxies in redshift bins of $\Delta z = 0.02$ between $0.15 \leq z \leq 0.49$. The bright end of the sample ($M_r \leq  -22.5$) is volume-limited in the whole redshift range. The low luminosity turn-offs are due to the limiting magnitude of $m_r \leq  19.5$ of the spectroscopic targeting algorithm of SDSS. The vertical dashed lines mark the magnitude interval in which galaxies are counted in order to scale the luminosity distribution curves together (see text). The bright ends of these curves run together reasonably well without any rescaling but correction for the small change in co-moving density is necessary to derive an overall luminosity function (cf. Fig.~\ref{fig:lumfscaled}).}
      \label{fig:lumf}
    \end{center}
\end{figure}

In Fig.~\ref{fig:lumfbyz} and \ref{fig:lumf} we plot the luminosity distribution of our spectroscopically selected sample in redshift bins of $\Delta z = 0.02$. The magnitudes are corrected for redshift and evolution. Because the co-moving density of early-type galaxies grows slightly with cosmic time the higher curves belong to lower redshifts. The cut-offs fainter than $M_r > -22.5$ are due to the magnitude limit of the sample, but we are only interested in the bright-end slope of these curves. Although the curves on Fig.~\ref{fig:lumf} run together reasonably well without any rescaling, in order to get an overall bright-end luminosity function that is valid over a broad redshift range, these curves need to be scaled together by compensating the density growth.

\begin{figure}
    \begin{center}
      \input{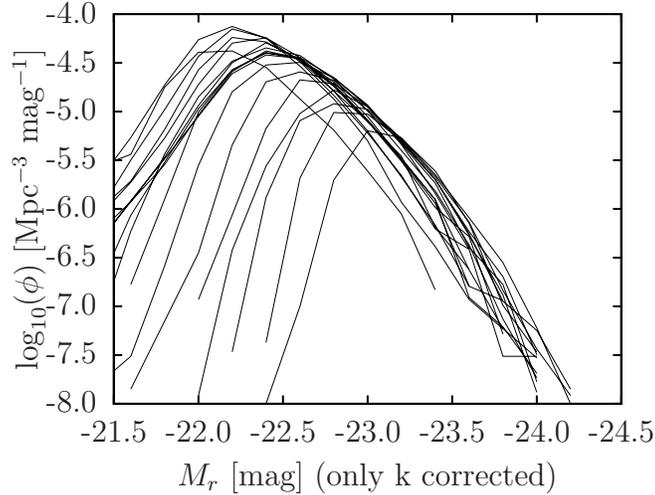}
      \caption{Luminosity distribution of the early-type galaxies in redshift bins of $\Delta z = 0.02$ between $0.15 \leq z \leq 0.49$. Only K-correction (and not evolution correction) was applied to the magnitudes (cf. Fig.~\ref{fig:lumf}).}
      \label{fig:lumfkcorr}
    \end{center}
\end{figure}

For reference, we also plot the luminosity distribution of the same galaxies with only K (but not evolution) corrected magnitudes in Fig.~\ref{fig:lumfkcorr}. Here curves do not overlap automatically.

\begin{figure}
	\begin{center}
		\input{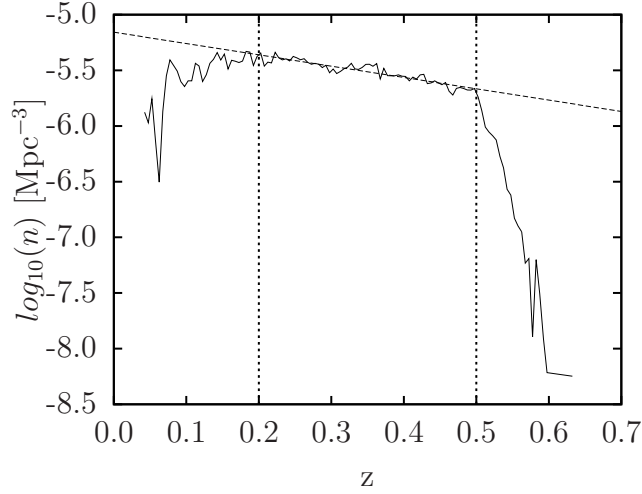}
		\caption{Co-moving density of the most luminous ($M_r \leq  -22.5$) early-type galaxies as a function of redshift. The number density decreases slowly with $z$ between $z \leq 0.15 \leq 0.5$. The turn-down at low redshift is thought to be due to the bright-end incompleteness of the sample, see Sec.~\ref{sec:spectro}. The fitting was done between $0.2 \leq z \leq 0.5$, as marked with the vertical lines.}
		\label{fig:density}
	\end{center}
\end{figure}

Fig.~\ref{fig:density} shows the change of the co-moving density of the brightest part of the early-type galaxy population. To get a volume-limited sample to as far as redshift $z = 0.5$, only galaxies having the r-band absolute magnitude between $-22.5 \geq  M_r \geq  -23.5$ are counted. The cut-off in density at $z = 0.5$ is due to the magnitude limit of the sample, while the cause of the cut-off at low redshift is explained in Sec.~\ref{sec:spectro}.

\begin{figure}
    \begin{center}
      \input{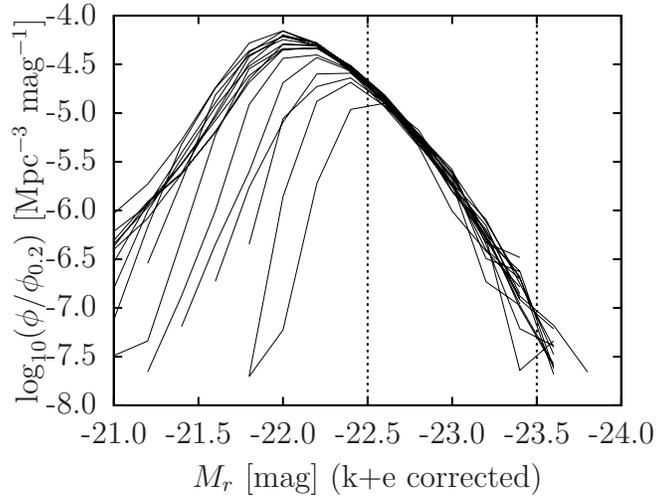}
      \caption{Luminosity distribution of the early-type galaxies in redshift bins of $\Delta z = 0.02$ between $0.15 \leq z \leq 0.49$ scaled to the density of galaxies brighter than $M_r < -22.5$ at $z = 0.2$. The scaling was applied based on the co-moving density evolution described in Sec.~\ref{sec:lumf} and plotted on Fig.~\ref{fig:density}. The curves now overlap very well in the $-22.5 > M_r > -23.0$ magnitude range. The scatter at higher luminosities is due to very low galaxy counts.}
      \label{fig:lumfscaled}
    \end{center}
  \end{figure}

The evolution of the co-moving number density only affects $\phi_0$. In the rest of the paper we used a normalized luminosity function that is independent of the galaxy density and show that the presented statistical method only depends on $\alpha$ and $M^*$. In order to scale the luminosity distribution curves of Fig.~\ref{fig:lumf} together and get Fig.~\ref{fig:lumfscaled}, we estimated the change of galaxy number density as a weak exponential function of redshift. We scaled the luminosity distributions of the individual redshift bins such a way that they all had the same co-moving density as the redshift bin at $z = 0.2$. The result of the fitting in the redshift range of $0.2 \leq 0.5$  is also plotted in in Fig.~\ref{fig:density}) with the slanted dashed line. The vertical lines represent the redshift range used for fitting. The fitted parameters of the function are the following.

\begin{equation}
n(z) = 10^{(-1.02 \pm 0.06) \cdot z + (-5.16 \pm 0.02)}

\end{equation}

This means a growth of about 0.3~dex in galaxy density from $z = 0.5$ to $z = 0.2$, which is consistent with the results of others (e.g. \citet{Bell2004, Faber2007}).

The results of the normalization is plotted in Fig.~\ref{fig:lumfscaled}. The curves run fairly well together at the bright end, but scatter at the very high luminosities is due to the very low number counts. The bright end of the luminosity distribution curves can be scaled together for our purposes because of their weak dependence on $\alpha$ and $M^*$, which are otherwise known to evolve slightly with redshift \citep{Bell2004, Faber2007}. These curves were averaged to get an overall luminosity distribution that is valid over a broad redshift range.

\begin{figure}
    \begin{center}
      \input{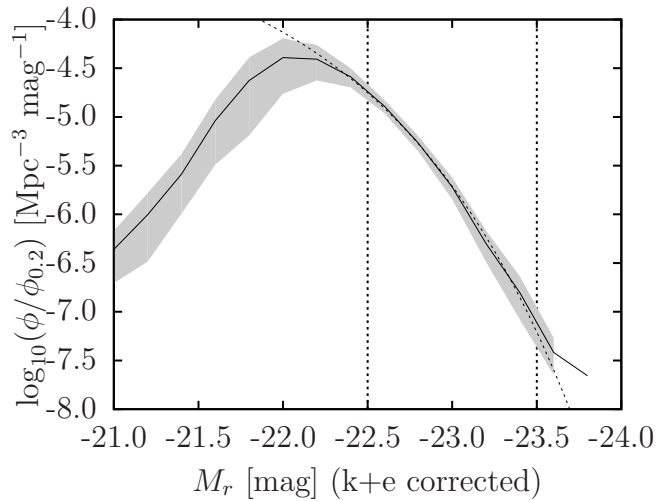}
      \caption{Luminosity distribution of the early-type galaxies normalized and averaged over redshift bins of $\Delta z = 0.02$ between $0.15 \leq z \leq 0.49$. The shaded area represents the standard deviation of the different curves plotted in Fig.~\ref{fig:lumfscaled}. The dashed line is the analytic fit to the data. The averaged distribution is a good representation of the magnitude distibution of the galaxies for $M_r < 22.5$. The sample is incomplete below this luminosity for higher redshift. Fitting with the Schechter function was done in the $-22.5 \geq M_r \geq -23.5$ range, as marked with the vertical lines.}
      \label{fig:lumfavg}
    \end{center}
  \end{figure}

Fig.~\ref{fig:lumfavg} shows the averaged luminosity distribution (solid line) and the fitted Schechter function in the form of Eq.~\ref{eq:schechter}. The fitting was done in the $-22.5 \leq M_r \leq -23.5$ range by keeping the faint-end parameter $\alpha$ fixed at $\alpha = -1.2$. This was done because $\alpha$ cannot be estimated sufficiently well from the brightest galaxies only. The resulting values of the parameters are $\alpha = -1.20 $, $\log_{10}{(\phi_0)} = -7.06 \pm 0.21 $, $M^* = -21.10 \pm 0.04$
 where $\phi_0$ is calculated with respect to $z = 0.2$. We will use this luminosity distribution in the following sections to give an introduction to order statistics.
  
\section{Extreme Value and Order Statistics}
\label{sec:evs}

\subsection{A brief introduction to EVS/OS}
\label{sec:evsintro}

Let us consider the probability density function $f(x)$. We are looking for the expectation value of the maximum (or minimum) of a sample of $N$ numbers drawn from this distribution randomly. The cumulative distribution function belonging to $f(x)$ is defined as usual:

\begin{equation}
	F(x) = \int_{-\infty }^x f(u) \, du
\end{equation}
hence, the probability of a number $x$ drawn from the distribution being less than a given $X$ is simply

\begin{equation}
	P(x < X) = F(X).
\end{equation}

Consequently, if we have a sample of $N$ independently drawn numbers $\{x_1, x_2, ... , x_N\}$ we can calculate the probability of $X_m = {\rm max} \{x_i\}$ being an upper limit of the sample as the probability of all numbers in the sample being less than $X_m$:

\begin{equation}
	P_m(X_m) = P(x_i < X_m) = P^N(x < X_m) = F^N(X_m).
\end{equation}
Differentiation with respect to $X_m$ yields the probability density function of the maximum of a sample of size of $N$:

\begin{equation}
	p_m(X_m, N) = N F^{N-1}(X_m) f(x),
\end{equation}
where $f(x)$ is the underlying (original) probability density function. All we have to do is to calculate the expectation value (or any higher moment) of $X_m$ drawn from $p_m(X_m, N)$ which, in case of a general $f(x)$ probability function and small $N$, is not possible analytically. It is very easy to calculate the integral numerically by the Monte-Carlo method, however.

Following the same pattern one can determine the probablity distribution, expectation value and moments of the $2^{nd}$, $3^{rd}$, etc.: the $k^{th}$ largest value of the sample. The probability distribution of the $k^{th}$ largest value is the following.

\begin{multline}
	p_{(k)}(X_{(k)}, N) = \\
	\frac{N!}{(k-1)!(N-k)!} (1- F(X_{(k)}))^{k-1} \, F^{N-k}(X_{(k)}) \, f(X_{(k)})
\end{multline}

The first theorem of extreme value statistics (Fisher-Tippet-Gnedenko theorem) states that in general the maximum (minimum) of a sample of $N$ numbers converges to one of the three possible distributions depending on the behaviour of the tail of the distribution from which the samples were drawn. This is considered the equivalent of the central limit theorem for extremes. Underlying distributions with exponential tails will all have the same extreme value distribution in the $N \rightarrow \infty $ limit (once normalized for the first two moments): the Gumbel distribution $g(x)$ \citep{FisherTippett1928, Gnedenko1948, Gumbel1958}.

\begin{equation}
	g(x) = e^{-x}e^{e^{-x}}
\end{equation}

It is important to emphasize that the Gumbel distribution is only applicable in the high-cardinality limit. Because the convergence is slow (usually logarithmic in $N$), and its speed also depends on the underlying distribution $f(x)$ at small sample sizes, significant corrections apply to the double exponential. These corrections also depend on $f(x)$ and analytic formulae are hard to calculate; numerical simulations are usually simple, however \citep{Gyorgyi2008}.

While other fields of science use extreme value statistics to calculate the probability of rare, catastrophic events (usually strong earthquakes, floods, stock exchange crashes etc.) in astronomy we do see these rare ``events'': the brightest objects of a certain kind. Because of the magnitude limit of the telescopes over a certain redshift only the most luminous galaxies, quasars, gamma ray bursts etc. are observable. Extreme value statistics might allow us to infer further information about the number of the undetected objects solely from the brightest detections if we have a model on the luminosity distribution of the objects.

\subsection{Applying EVS/OS to the early-type galaxy luminosity function}
\label{sec:evsprop}

In this section we show an example on how EVS/OS is applicable to extrapolate early-type galaxy counts beyond the redshift where the magnitude limit of the spectroscopic sample starts to affect the measured galaxy counts.

Although the Schechter function, if written as a function of the magnitude, is not a valid probability density as it is not capable of being integrated over all magnitudes, it becomes integrable and can be normalized whenever a lower magnitude limit is introduced.

We took the analytic fit to the measured, rescaled and averaged $r$-band luminosity distribution (as explained in Sec.~\ref{sec:lumf} and plotted on Fig.~\ref{fig:lumfavg}), applied a cut-off of $M_r < -22$~mag, and ran the Monte-Carlo integration to get the expectation value of the maximum luminosity (minimum magnitude) as a function of the sample size $N$, and also the expectation values of the $2^{nd}$, $5^{th}$, $10^{th}$, $20^{th}$ and $100^{th}$ most luminous galaxy. The results are plotted in Fig.~\ref{fig:evs}. As the plot shows, the expectation value of the luminosity raises rapidly with $N$ for small values of $N$ and turns into a slowly raising curve at higher sample size. For the Schechter parent distribution, these curves take the generic form of $\ln{\ln{N}}$ for large enough $N$ because around the expectation value of the maximum $L_m$ the following is true.

\begin{equation}
	N \, \delta \left( \frac{L_m}{L^*} \right)^{\alpha} \exp{\left(- \frac{L_m}{L^*} \right)} \approx 1,
\end{equation}
where $\delta$ is the unit interval on the scale of $L/L^*$. For $L_m \gg L^*$ this yields

\begin{equation}
	\frac{L_m}{L^*} \approx \ln{N} + O(\ln{\ln{N}}).
\end{equation}

The second logarithm comes from the logarithm of the magnitude scale. The very slowly rising behaviour of these $\ln{\ln{N}}$ curves is the reason behind the fact that the brightest cluster galaxies can be used as standard candles (for example \citet{Sandage1972, Sandage1976, Postman1995, Whiley2008}). (Here we assume for a moment that BCGs are drawn from the same distribution as the rest of early-type galaxies. We will discuss this issue later.) The expectation value of the luminosity is evidently decreases with $k$, the dependence is also logarithmic.

\begin{figure}
    \begin{center}
      \input{figures/evs.tex}
      \caption{Expectation value of the absolute magnitude $M_{(k)}$ of the $1^{st}$, $2^{nd}$, $5^{th}$, $10^{th}$, $20^{th}$ and $100^{th}$ (from top to down) most luminous LRG (brighter than $M_r < -22$) as a function of the sample size $N$, computed directly from the luminosity function using the Monte-Carlo method, averaged over 5,000 runs. The vertical dashed lines mark the sample sizes of $N = 100$, $200$, $500$ and $1,000$ which are used to plot Fig.~\ref{fig:evsvar} and Fig.~\ref{fig:evsavg}. The curves show a $\log{\log{N}}$ behaviour.}
      \label{fig:evs}
    \end{center}
  \end{figure}
  
The standard deviation of $M_{(k)}$ shows a more interesting behaviour that limits the usability of EVS/OS for high-precision studies and also explains some well-known rules of thumb of astronomy. In Fig.~\ref{fig:evssigma} we plot the standard deviation of the expectation value of the magnitude of the $1^{st}$, $2^{nd}$, $5^{th}$, $10^{th}$, $20^{th}$ and $100^{th}$ most luminous galaxy as a function of the sample size $N$. By looking at the individual curves, it is clear that the variance decreases only slightly with bigger sample size $N$, but changes significantly with higher order $k$. This is the reason that, for example, the magnitude of the brightest cluster galaxies changes more from cluster to cluster than the magnitude of the second, fifth etc. brightest galaxy and why they are better standard candles than BCGs \citep{Scott1957}.
  
\begin{figure}
    \begin{center}
      \input{figures/evs_sigma.tex}
      \caption{Standard deviation of the absolute magnitude $M_{(k)}$ of the $1^{st}$, $2^{nd}$, $5^{th}$, $10^{th}$, $20^{th}$ and $100^{th}$ (from top to down) most luminous LRGs (brighter than $M_r < -22$) as a function of the sample size $N$, computed directly from the luminosity function using the Monte-Carlo method, averaged over 5,000 runs. The vertical dashed lines mark the sample sizes of $N = 100$, $200$, $500$ and $1,000$, which where used to plot Fig.~\ref{fig:evsvar} and Fig.~\ref{fig:evsavg}. The curves change only slightly with $N$ for $N > 100$.}
      \label{fig:evssigma}
    \end{center}
  \end{figure}
  
\begin{figure}
    \begin{center}
      \input{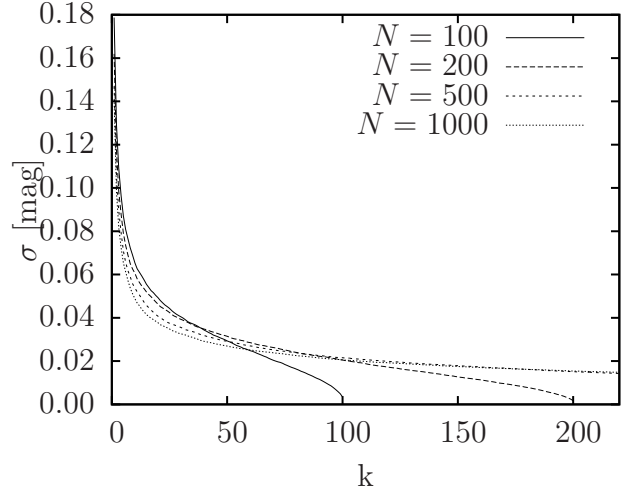}
      \caption{Standard deviation of $M_{(k)}$ in a random sample of $N = 100$, $200$, $500$ and $1000$ galaxies as a function of $k$. An inflexion of the curves is observable around $k \simeq  N/2$; above this, the variance does not change significantly with $k$.}
      \label{fig:evsvar}
	\end{center}
\end{figure}

In Fig.~\ref{fig:evsvar}, we plot the standard deviation of $M_{(k)}$ as a function of $k$ for four different $N$. As the plot shows, the variance decreases very rapidly with $k$ and the curves have an inflexion point around $k \simeq  N / 2$. Above $k \simeq N/2$, the variance decreases only mildly with $k$. This means that if we are able to measure the apparent magnitude and redshift of 5 -- 10~\% of the most luminous objects in a certain area of the sky or in a certain redshift bin, we can find standard candles with an accuracy of 0.2~mag. The actual directly measured number counts in our redshift bins at higher redshifts are around 1000. Consequently, if we want to apply EVS/OS techniques to determine the number count of the fainter galaxies successfully in a certain redshift range, it would be necessary to measure the spectroscopic redshift of the brightest 50 -- 100 galaxies per redshift bin. This targeting process requires accurate photo-z and absolute-magnitude estimates.

\begin{figure}
    \begin{center}
      \input{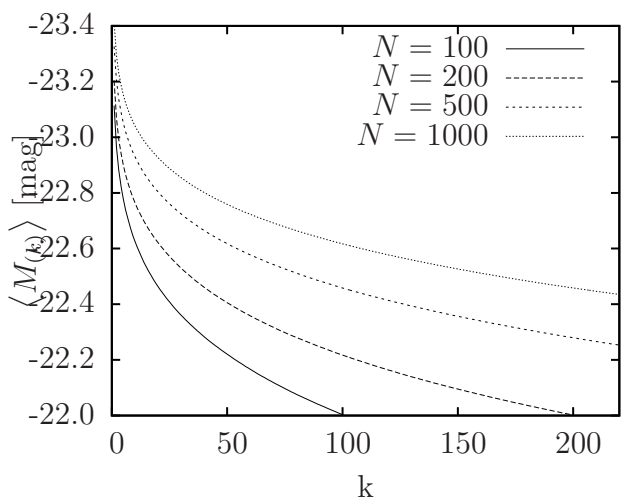}
      \caption{Expectation value of $M_{(k)}$ in a random sample of $N = 100$, $200$, $500$ and $1000$ galaxies brighter than $M_r < -22$. }
      \label{fig:evsavg}
    \end{center}
  \end{figure}

For reference, we also plot the expectation value of $M_{(k)}$ as a function of $k$ for four different $N$ in Fig.~\ref{fig:evsavg}. The important thing to observe here is that the curves actually reach the magnitude limit of $M_r = -22$ in the limit of $k \rightarrow  N$ (only visible for $N = 100$ and $N = 200$). This is only true for underlying probability density functions with a hard cut-off.

\begin{figure}
    \begin{center}
      \input{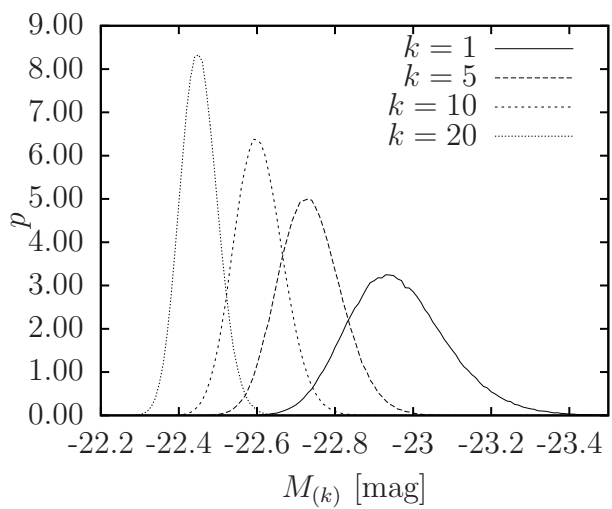}
      \caption{Probability distribution of the magnitude of the first, fifth, tenth and twentieth most luminous galaxy in random samples of cardinality $N = 100$. Results from a Monte-Carlo simulation of 500,000 runs.}
      \label{fig:evsgumbel}
    \end{center}
  \end{figure}

In Fig.~\ref{fig:evsgumbel} we plot the distribution of the magnitude of the $k^{th}$ most luminous galaxies drawn from a distribution based on the early-type luminosity function for $k = 1$, $5$, $10$ and $20$. The sample size was fixed at $N = 100$ and the simulation was run 500,000 times. The curves show distributions typical to order statistics with slightly asymmetric probability density functions. For $k = 1$ and $N \rightarrow \infty $ the distribution should be the Gumbel distribution, for $k \rightarrow \infty $ and $N \rightarrow \infty $ the limiting distribution is Gaussian. The actual $k = 1$ cannot be fitted very well with the theoretical Gumbel distribution because of the small $N$, as discussed in Sec.~\ref{sec:evsintro}: convergence to the Gumbel distribution is slow \citep{Gyorgyi2008, Manu2010}. The FWHM of the $k = 1$ curve is consistent with what we know about BCGs, i.e. the scatter in their absolute magnitude is around $\Delta M \simeq 0.25 \,\, \text{mag}$ \citep{Sandage1972, Sandage1988, AragonSalamanca1998} from cluster to cluster.
  
\section{Estimating galaxy counts from the brightest LRGs}
\label{sec:counts}

\begin{figure*}
  \begin{minipage}{1\textwidth}
    \begin{center}
      \input{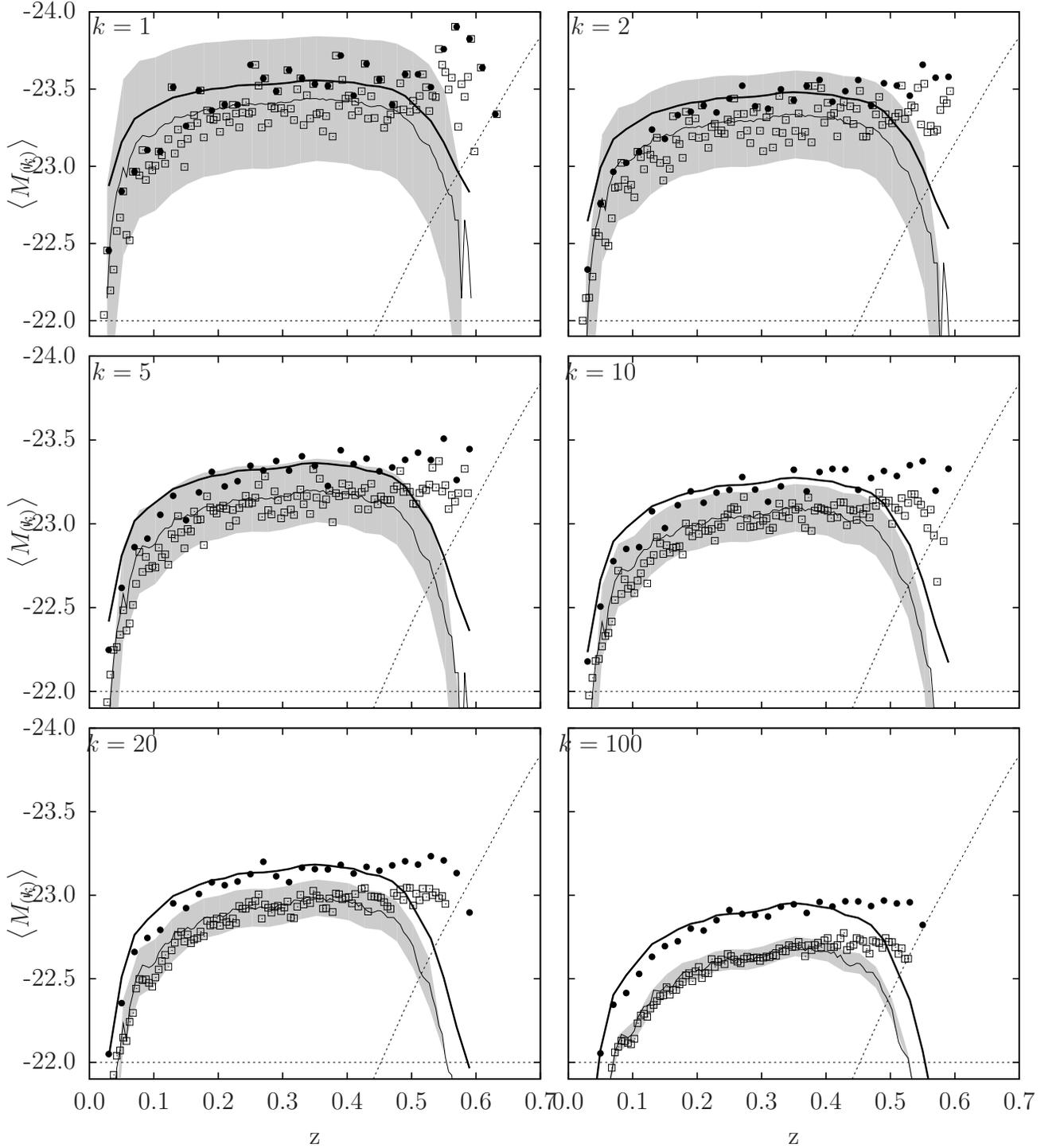}
      \caption{Absolute magnitudes of the first, second, fifth, tenth, twentieth and hundredth most luminous galaxy, as functions of redshift in redshift bins of $\Delta z=0.02$ (black circles) and $\Delta z=0.005$ (open rectangles). The solid curves show the expectation value of $M_{(k)}$ as calculated from the measured galaxy counts in the redshift bin based on extreme value statistics. The thick lines represent magnitudes calculated for the big redshift bins, while the thin lines are for the small bins. Note that fluctuations of the measured magnitudes get smaller with higher $k$. The horizontal dashed line represents the magnitude limit of galaxy counts (and also the validity of the statistical models) while the slanted dashed line is the equivalent of the apparent magnitude limit of the sample selection. The sample becomes incomplete at the redshift where these two lines intersect, this is where the magnitude estimates based on galaxy counts diverge from the measurements.}
      \label{fig:maxlum}
    \end{center}
    \end{minipage}
  \end{figure*}

To confirm our assumption that order statistics can be used to accurately estimate the brightest galaxies' absolute magnitude solely from number counts based on a single, non-evolving luminosity function, we refer to Fig.~\ref{fig:maxlum}. Here we plot the magnitude of the $k^{th}$ brightest galaxy in each redshift bin for two different bin sizes: big -- $\Delta z = 0.02$, and small -- $\Delta z = 0.005$. The black circles represent the brightest magnitudes of the big bins (thus they are usually brighter than in the small bins) while the open rectangles represent the brightest magnitudes of the small bins. The curves represent the expectation value of $M_{(k)}$ as computed from the number counts in the redshift bins (thick line -- big bins, thin line -- small bins) based on order statistics. For the small bins, we also plot the $2\sigma$ range (shaded area arount the thin curve). The curves computed from number counts agree reasonably well with the actual luminosity measurements in the redshift range where all galaxies brighter than $M_r < -22$ were observed.

\begin{figure*}
  \begin{minipage}{1\textwidth}
    \begin{center}
      \input{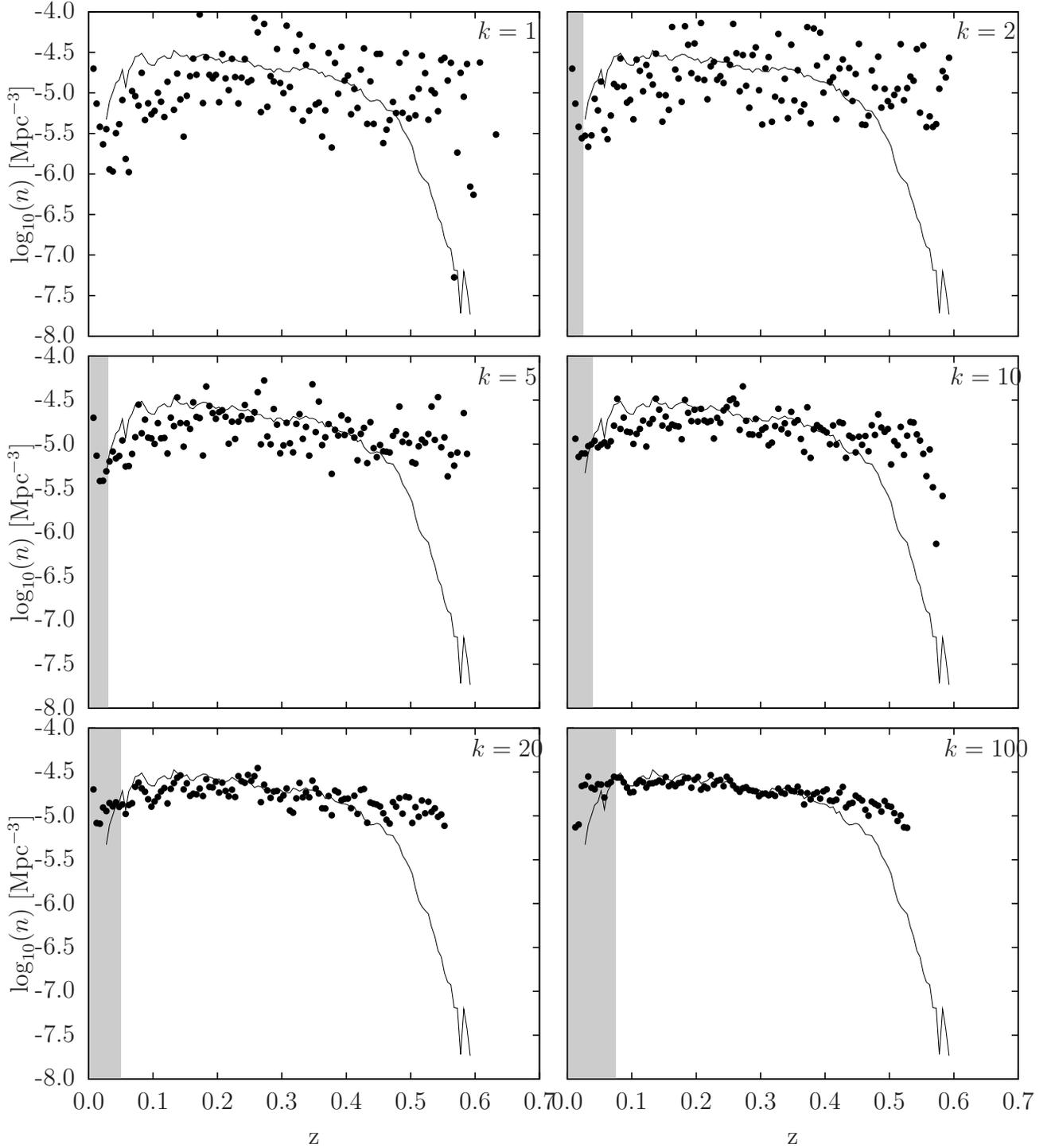}
      \caption{Number density of galaxies brighter than $M_r < -22$ in redshift bins of $\Delta z = 0.005$. The points represent the predictions made from the absolute magnitude of the first, second, fifth, tenth, twentieth and hundredth most luminous galaxy of the redshift bin based on extreme value statistics. Solid lines (same on all six plots) are direct measurements of the galaxy density. Note that estimation from brighter galaxies allows extrapolation to higher redshifts, but the noise is significantly higher compared to higher orders. The shaded areas represent regions where the luminosity of the $k^{th}$ brightest galaxy is below $M_r > -22$. In this range, the models are extrapolations only, and are not necessarily valid.}
      \label{fig:counts_norm}
    \end{center}
    \end{minipage}
  \end{figure*}

By inverting the functions plotted in Fig.~\ref{fig:evs}, one can easily calculate the galaxy counts in a redshift bin from the magnitude of the $k^{th}$ brightest galaxy of the bin. Fig.~\ref{fig:counts_norm} shows our estimate of galaxy density for six different values of $k$ as a function of $z$. For low values of $k$ the estimates show very large scatter, but the fluctuations of the estimate for $k = 100$ are already comparable to the fluctuations of the directly measured galaxy density. These plots also show the extrapolation capabilities of the method. Directly measured galaxy counts go down around $z \simeq 0.45$ because of the apparent magnitude limit of spectroscopic target selection, but the brightest galaxies are still in the sample up to redshift $z = 0.55$ -- $0.6$, and allow us to calculate galaxy counts for this redshift range too.
  
\begin{figure*}
  \begin{minipage}{1\textwidth}
    \begin{center}
      \input{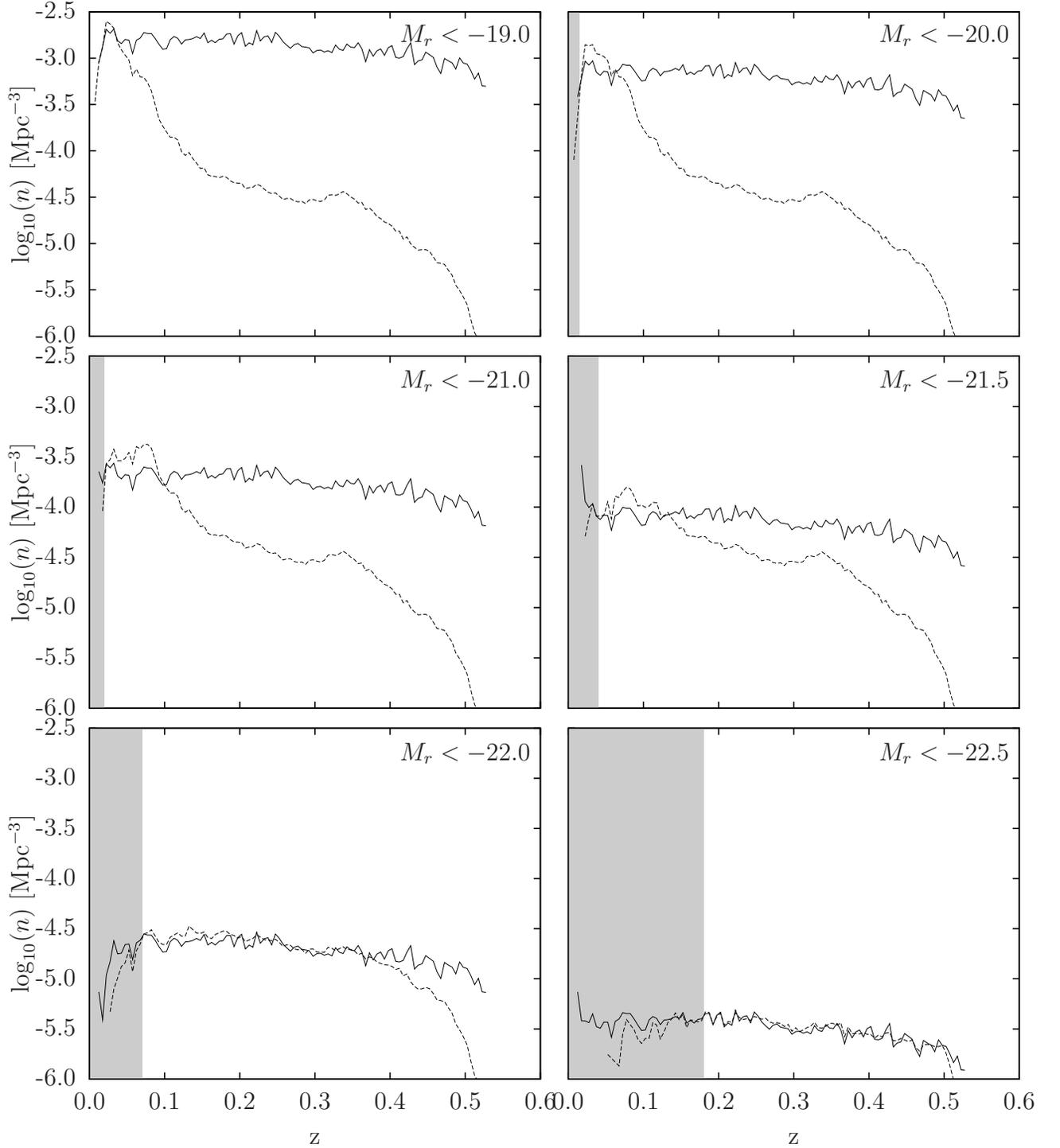}
      \caption{Extrapolation capabilities of order statistics. The plots show the co-moving number density of LRGs as a function of redshift calculated in redshift bins of $\Delta z = 0.005$. Different plots correspond to different absolute magnitude cuts in the SDSS $r$ band. Dashed lines represent the galaxy density directly calculated from galaxy counts, while the solid lines show the order statistics estimates from the magnitude of the $100^{th}$ brightest galaxy of the redshift bins. The shaded areas show the redshift regions where the magnitude of the $100^{th}$ brightest galaxy is below the magnitude cut we used to calculate number counts; here, OS estimates are invalid. It is clear from the plots that at low magnitude cuts, the selection effects dominate the sample, and directly measured galaxy counts are different from the estimates based on the brightest magnitudes. At low redshift, the contamination of the sample is also observable. The turn-down of the dashed curves at high redshift is due to the magnitude limit of the \texttt{spectro} sample. }
      \label{fig:extrapolate}
    \end{center}
    \end{minipage}
  \end{figure*}

We emphasize the extrapolation capabilities of our method to redshift ranges where the magnitude limit of the telescope or the target selection criteria forbid computing galaxy density directly in Fig.~\ref{fig:extrapolate}. The plots show OS estimates on the galaxy number density for six different absolute magnitude cuts from the brightness of the $100^{th}$ most luminous galaxy in redshift bins of $\Delta z = 0.005$. For low magnitude cuts, the target selection effects evidently determine the directly measured number counts. For brighter cuts, the OS estimates nicely converge to the measurements. It is important to note that we used the same luminosity function to calculate the OS quantities; only the integration limit was changed.
   
\section{Statistics of the BCGs}
\label{sec:bcg}

It has been thought for decades \citep{Dressler1976, Ostriker1977}, recently investigated by \citet{Loh2006, Lin2010} that the brightest central galaxies of the clusters cannot be explained as extreme values drawn from the \em same \em luminosity distribution as the rest of the galaxies in rich clusters. The main argument was that the observed magnitude gap of $M_{12} \simeq 0.8 \,\, \text{mag}$ between the brightest and the second brightest member of a galaxy cluster cannot be explained if both galaxies are drawn \em randomly \em from the same underlying luminosity distribution. On the other hand, as we demonstrated in Sec.~\ref{sec:evsprop} and Sec.~\ref{sec:counts}, this is not necessarily the case if we look at the whole LRG population without considering galaxies to be part of clusters. The absolute magnitudes and the scatter of the magnitudes of extremely luminous red galaxies seem to be well-described by the single Schechter function that can be derived from the total ensemble of LRGs once the correct statistical method is applied.

As we demonstrated in Sec.~\ref{sec:evsprop}, the expectation value of the magnitude difference $M_{12} = M_{(1)} - M_{(2)}$ between the brightest and the second-brightest galaxy of a \em random sample \em drawn from the LRG luminosity function is well below $\left<M_{12}\right> \leq  0.2 \,\,\text{mag}$, cf. Fig.~\ref{fig:evs}.

\begin{figure}
    \begin{center}
      \input{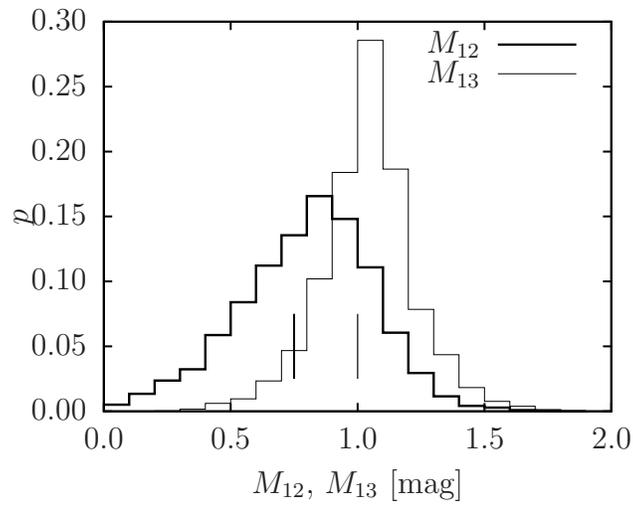}
      \caption{Probability distribution of the magnitude difference of the first and the second and third most luminous galaxy ($M_{12}$ -- thick line, $M_{13}$ -- thin line) in a sphere with a radius of $r = 10\,\,\text{Mpc}$ around simulated BCGs. The small vertical bars mark the expectation values. Results from a Monte-Carlo simulation of 10,000 runs.}
      \label{fig:bcghist}
    \end{center}
  \end{figure}

In order to resolve this contradiction we investigated the statistics of the following very simple model. We populated uniformly a $1 \,\, \text{Gpc}^3$ volume with galaxies at a number density of $10^{-4.5} \,\,\text{Mpc}^{-3}$. Next we assigned absolute magnitudes brighter than $M < -22$ to all of these galaxies. The magnitudes were randomly drawn from the LRG luminosity function we determined previously. We considered every galaxy a BCG if it had the absolute magnitude brighter than $M < -23$. For every such BCG, we determined the magnitude of the second and third brightest galaxy within $10 \,\,\text{Mpc}$. The distribution of the magnitude difference between the BCGs and the second and third brightest galaxies is plotted in Fig.~\ref{fig:bcghist}. The expectation value of the gaps are $M_{12} \simeq 0.75$ and $M_{13} \simeq 1.0$, in good agreement with the observations.

The model is very primitive compared to our knowledge about the large-scale structure of the universe, but one can consider the randomly positioned galaxies of the real space as the centers of the main progenitor DM halos of future massive galaxies at some very early age of the Universe when the mass distribution was homogeneous on small scales. The $10 \,\,\text{Mpc}$ cluster radius can be imagined as the radius of the DM halos from which the central region of a future cluster will form. Of course, the final cluster will have a much smaller radius due to the merging of these halos.

\section{Discussion and summary}
\label{sec:discussion}

We have shown in Sec.~\ref{sec:counts} and in Figs.~\ref{fig:maxlum}-\ref{fig:extrapolate} that EVS/OS can be successfully applied to the LRG sample of SDSS and yield consistent magnitudes and galaxy counts, with significant extrapolation capabilities if selection effects dominate the sample. Since the SDSS data set covers a significant part of the sky, we evidently included galaxy clusters of all sizes in the sample. Consequently, the sample contains all BCGs of these clusters. This raises two important questions that we would like to discuss in this section.

\subsection{The effect of clustering}

In order to demonstrate the extrapolation capabilities of the presented technique, we chose to bin our data in redshift bins instead of spatial bins, and all of these bins contained very bright cluster galaxies, because SDSS covers a very large area on the sky. What would happen if we restricted our sample to areas of the sky where only fainter galaxies are present? From fainter maximal luminosities, order statistics would estimate galaxy counts to be accordingly lower. A fainter sample (in terms of absolute magnitudes) means fewer galaxies since the sample was probably targeted to avoid dense clusters. On the other hand, samples targeting brighter galaxies would cover dense groups and clusters with lots of galaxies; consequently, galaxy counts would be higher. EVS/OS can only be applied to samples covering a big enough area of the sky to cancel the bias caused by clustering.

\subsection{Brightest cluster galaxies}

Our resolution to the controversy about the magnitude gap experienced in case if BCGs is that clusters \em cannot be considered random samples \em in which galaxies are drawn independently from any distribution. Samples containing a BCG are special. We have showed in Sec.~\ref{sec:bcg} that if spatial distribution of galaxies is also taken into account, one can very easily reproduce the measured statistics of clusters solely on the basis of a single LRG luminosity function. This result suggests that BCGs are indeed formed the same way as the rest of the LRGs, and the scenario of galactic cannibalism can be ruled out. However, our model is extremely simple and the statistics of BCGs requires more investigation based on N-body simulations and merger trees. We will publish our results in a future paper \citep{Dobos2011}.

\subsection{Possible application of order statistics based estimations in future surveys}

The demonstrated statistical methods highly depend on accurate absolute-magnitude measurements and spectroscopic redshifts. Upcoming large surveys like PanSTARRS and LSST will provide deep photometric datasets without any directly measured redshift information. The intrinsic error of photometric redshift methods is about $\Delta z \simeq 0.01$ -- $0.02$ \citep{Hildebrandt2010}. This is much larger than our redshift bin size, so redshift measurements with significantly better accuracy are required to apply our method with the same resolution. In light of extreme-value and order statistics, besides aiming for a complete spectroscopic follow-up survey of the large photometric surveys, it might be also useful to calculate preliminary redshifts based on photometry and target the brightest galaxies \em per redshift bin \em first for spectroscopy. This would reduce the number of targets significantly (a couple of thousands instead of hundreds of thousands), and still galaxy counts could be calculated at very far distances, even where only the brightest galaxies are observable.

The $k^{th}$ brightest galaxies (for large values of $k$) are useful standard candles which can be used to test cosmological models. However, in case of early-type galaxies, this method is limited by the quality of the evolutionary model used to compute absolute magnitudes, just as by the model of the evolution of $\alpha$ and $M^*$ of their luminosity function.

\section*{Acknowledgments}

The authors would like to thank Zolt\'an R\'acz for his very useful help in the field of extreme value statistics.

This work was supported by the following Hungarian grants: NKTH: Pol\'anyi, OTKA-80177 and KCKHA005.

Funding for the SDSS and SDSS-II has been provided by the Alfred P. Sloan Foundation, the Participating Institutions, the National Science Foundation, the U.S. Department of Energy, the National Aeronautics and Space Administration, the Japanese Monbukagakusho, the Max Planck Society, and the Higher Education Funding Council for England. The SDSS Web Site is http://www.sdss.org/.

The SDSS is managed by the Astrophysical Research Consortium for the Participating Institutions. The Participating Institutions are the American Museum of Natural History, Astrophysical Institute Potsdam, University of Basel, University of Cambridge, CaseWestern Reserve University, University of Chicago, Drexel University, Fermilab, the Institute for Advanced Study, the Japan Participation Group, Johns Hopkins University, the Joint Institute for Nuclear Astrophysics, the Kavli Institute for Particle Astrophysics and Cosmology, the Korean Scientist Group, the Chinese Academy of Sciences (LAMOST), Los Alamos National Laboratory, the Max-Planck-Institute for Astronomy (MPIA), the Max-Planck-Institute for Astrophysics (MPA), New Mexico State University, Ohio State University, University of Pittsburgh, University of Portsmouth, Princeton University, the United States Naval Observatory, and the University of Washington.


\label{lastpage}

\end{document}